\documentclass[12pt]{article}
\usepackage{epsfig}
\usepackage{amsmath}
\usepackage{amssymb}

\begin{document}

\title{Production of $\omega$-mesons in proton-proton collisions}

\author{K. Nakayama$^{a,b}$, A. Szczurek$^c$, C. Hanhart$^a$, \\
J. Haidenbauer$^a$, and J. Speth$^a$ \\ \\
{\small $^a$Institut f\"{u}r Kernphysik, Forschungszentrum J\"{u}lich
GmbH,}\\ {\small D--52425 J\"{u}lich, Germany} \\
{\small $^b$Department of Physics and Astronomy, University of Georgia,}\\{\small  Athens, GA 
30602, USA} \\
{\small $^c$Institute of Nuclear Physics,}\\{\small PL-31-342 Cracow, Poland}
}

\maketitle

\begin{abstract}
The production of $\omega$-mesons in proton-proton collisions for proton
incident energies up to $2.2$ GeV is investigated within a meson-exchange
model of hadronic interactions. We find a large cancellation between the 
dominant $\pi\rho\omega$ meson-exchange current and nucleonic current 
contributions. A comparison with preliminary data from SATURNE calls for the 
inclusion of off-shell form factors at the $NN\omega$ and $\pi\rho\omega$ 
production vertices. Due to the present lack of knowledge of these form 
factors, together with the destructive interference mentioned above, the 
relative magnitude of the nucleonic and meson-exchange current contributions 
cannot be determined from existing total cross section data. However, it is 
shown that the angular distribution of the produced $\omega$-mesons provides 
an unique and clear signature of the magnitude of these currents, thus 
allowing one to disentangle these two basic reaction mechanisms.
\end{abstract}

\newpage

 \section{Introduction}

Vector-meson production in both hadronic and electromagnetic processes are 
considered to be an excellent tool for investigating the properties of vector 
mesons, both in medium and in free space. The properties of these mesons in 
the nuclear medium appear to be of special interest not only for the 
understanding of the nuclear dynamics but also for possibly revealing 
information about the deconfinement phase transition of hadrons to the 
quark-gluon plasma or the restoration of chiral symmetry at high baryon 
density and/or temperature \cite{Shor,Brow,Hats,Asak}. The investigation of 
vector meson production processes in free space is important too. In addition 
to providing the necessary elementary production amplitudes required for
in-medium studies, one can address other basic questions. For example, the
nucleon-nucleon-vector-meson ($NNv$) vertex functions, even on the mass-shell
(coupling constants), are largely unknown, especially for the $\omega$ and
$\phi$ mesons. Vector-meson production in proton-proton ($pp$) collisions may
offer a means of extracting these coupling constants. Specifically, in the 
case of the $\phi$-meson production, one might be able to obtain information 
on the strangeness content of the nucleon.

Therefore it might be surprising that very little information about 
vector-meson production processes can be found in the literature. Only quite 
recently have these reactions begun to receive increasing attention. For 
example, the near-to-threshold production of $\omega$-mesons in the reaction 
$p + d \rightarrow {^3}$He$ + \omega$ has been investigated both 
experimentally \cite{Wurz} and theoretically \cite{Wilkin,Kond}. Also the 
production of both the $\omega$- and $\phi$-mesons in $pp$ collisions close to
threshold is now being analysed by the SPES3 collaboration at SATURNE (Saclay)
\cite{Saclay} and will soon be studied at COSY (J\"ulich) \cite{COSY}. From 
the theoretical side, Sibirtsev \cite{Sib96} has investigated the $\rho$-, 
$\omega$- and $\phi$-meson production within a simple one-pion exchange model,
utilizing parametrizations of the measured cross sections in 
$\pi N \rightarrow  v N$ ($v = \rho ,\omega ,\phi$) processes. Chung et al. 
\cite{CLK97} have calculated the $\phi$-meson production in $\pi$-baryon and 
baryon-baryon collisions within the meson-exchange picture.

Clearly these works aim mainly at a parametrization of the 
$N + N \rightarrow N + N + v$ reactions which is suitable for application in 
investigations of vector-meson production in proton-nucleus and heavy-ion 
collisions \cite{Gol93,Sib97,Chung97}, and not so much at a detailed analysis 
of these reactions themselves. In particular, in none of these calculations is
the $NN$ final state interaction (FSI) taken into account. In particle 
production reactions near their kinematical thresholds the final $NN$ subsystem
emerges at fairly low energies. Therefore, the nucleons are mainly in S-wave 
states where their interaction is very strong.  Indeed, large effects of the 
$NN$ FSI are well-known from other processes such as the $N + N \rightarrow 
N + N + \gamma$ \cite{Kanzo} and $N + N \rightarrow N + N + \pi$ reactions 
\cite{Meyer}.

In the present work we focus on the $p + p \rightarrow p + p + \omega$ process.
The reason for studying $\omega$-meson production is that, in contrast to 
other vector-meson production reactions, only a few relevant basic production 
mechanisms are involved, and consequently, it is the simplest reaction among 
the various vector-meson productions in $pp$ collisions.  We describe this 
reaction within a meson-exchange model which takes the $NN$ FSI into account. 
Our model calculation includes both the nucleonic and meson-exchange currents 
as defined below. The initial state interaction (ISI) is neglected. In this 
regard we mention that, quite recently, M. Batini\'c at el. \cite{BSL} has 
examined its influence for the case of the reaction $p + p \rightarrow 
p + p + \eta$. These authors found that the ISI leads only to an overall 
reduction of the total cross section (by about 40\%) but has virtually no 
effect on its energy dependence. Based on this result we would not expect any 
significant influence of the ISI on the qualitative aspects of $\omega$ 
production discussed in the present work.

  The paper is structured as follows: In Section II we describe the details
of our model. We give the basic formalism and specify the parameters
entering into the numerical computations. The reults of our calculations are
presented in Section III where we show that the nucleonic current as well as 
the $\pi \rho \omega$ meson-exchange current yield potentially large 
contributions to the $\omega $-production cross section. The inclusion of form
factors at the $\omega$-production vertices is necessary if one wants to 
achieve results comparable (in magnitude) to the preliminary data from 
SATURNE. We also find a strong cancellation between the contributions from the
nucleonic current and meson-exchange currents - which makes it rather 
difficult to discriminate between these two production mechanisms from a study
of the total cross section alone. Fortunately, as we shall demonstrate, the 
angular distribution of the produced $\omega$-mesons shows a clear dependence 
on the magnitude of these currents and is therefore suitable for disentangling
the two dominant reaction mechanisms in question. A summary of our results is 
given in Section IV.

 \section{Details of the model}

\begin{figure}
\epsfig{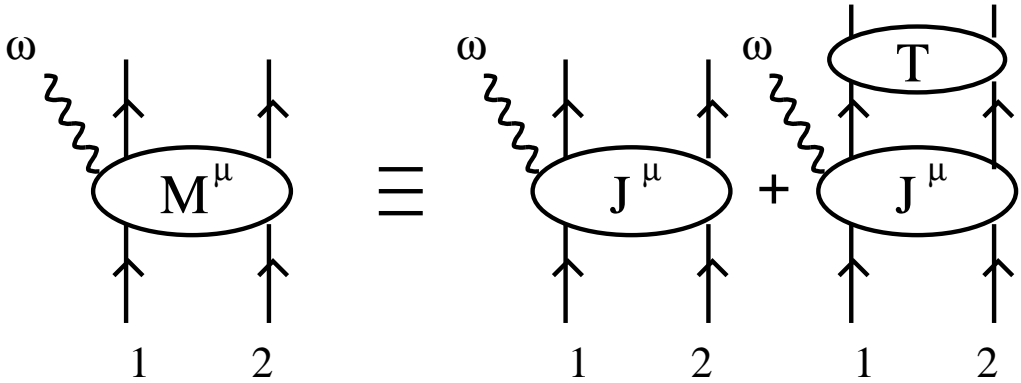}
\caption{The transition amplitude for the reaction $p + p \rightarrow 
p + p + \omega$.}
\label{fg1}
\end{figure}

We write the transition amplitude describing the
$p + p \rightarrow p + p + \omega$ process as
\begin{equation}
M^\mu = < \phi_f | (T^{(-)\dagger}_f G_f + 1) J^\mu | \phi_i > \ ,
\label{amplitude}
\end{equation}
where $\phi_{i,f}$ denotes the four-component unperturbed $NN$ wave function
in the initial ($i$) and final ($f$) state. $T^{(-)}_f$ is the final state
$NN$ T-matrix. $G_f$ stands for the two-nucleon propagator and $J^\mu$ is the
$\omega$-emission current. With the $\omega$-emission current (defined below) 
taken only in Born approximation, the above transition amplitude corresponds to
a Distorted Wave Born Approximation with no ISI. Eq.(\ref{amplitude}) is 
diagrammatically depicted in Fig.~\ref{fg1}.  

Before discussing the structure of $J^\mu$ in detail let us first specify the 
$NN$ T-matrix. The T-matrix used in our calculation is generated by solving a 
three-dimensional reduced Bethe-Salpeter equation (the Blankenbecler-Sugar
equation) for a relativistic one-boson-exchange $NN$ potential $V$, i.e.,
\begin{equation}
T = V + ViG_{BBS}T \ ,
\label{scateqn}
\end{equation}
where $G_{BBS}$ denotes the Blankenbecler-Sugar(BBS) two-nucleon propagator.
In this work we employ the Bonn B $NN$ model as defined in table A.1 of
Ref.\cite{Mach89}, which includes the $\pi$-, $\rho$-, $\omega$-, $\sigma$-, 
$\eta$-, and $a_o$-mesons. This interaction model fits the $NN$ phase-shifts up
to the pion threshold energy as well as the $NN$ low-energy parameters and the
deuteron properties \cite{Mach89}. It should also be mentioned that each
nucleon-nucleon-meson ($NNM$) vertex in the $NN$ potential is supplemented by a
form factor either of monopole or dipole form. We refer to \cite{Mach89} for 
further details.  Furthermore we note that the two-nucleon propagator $G_f$ 
appearing in Eq.(\ref{amplitude}) is, for consistency, also chosen to be the 
BBS propagator, $G_f = G_{f;BBS}$.

\begin{figure}
\epsfig{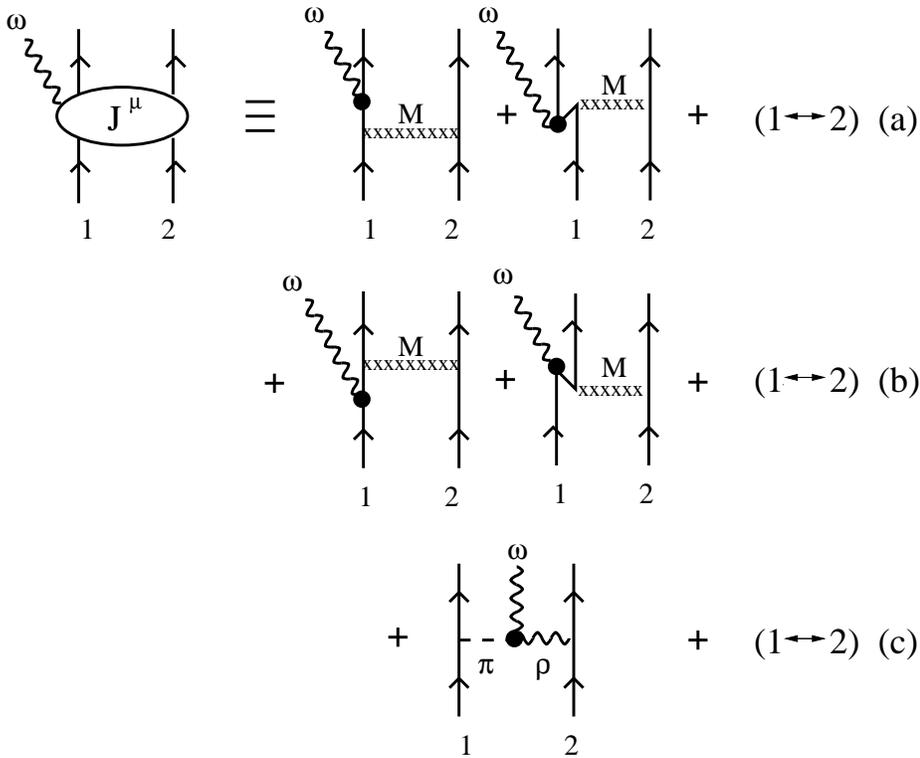}
\caption{$\omega$-meson production currents, $J^\mu$, included in the present 
study. (a) and (b) are the nucleonic current, and (c) is the meson exchange 
current. $M = \pi, \eta, \rho, \omega, \sigma, a_o$.}
\label{fg2}
\end{figure}

The $\omega$-emission current $J^\mu$ in Eq.(\ref{amplitude}) is given by the
sum of the nucleonic and meson-exchange currents, $J^\mu = J^\mu_{nuc}  + 
J^\mu_{mec}$, as illustrated diagrammatically in Fig.~\ref{fg2}. The nucleonic
current is defined as
\begin{equation}
J^\mu_{nuc} = \sum_{j=1,2}\left ( \Gamma^\mu_j iS_j U + U iS_j \Gamma^\mu_j 
\right ) \ ,
\label{nuc_cur}
\end{equation}
with $\Gamma^\mu_j$ denoting the $NN\omega$ vertex and $S_j$ the nucleon
(Feynman) propagator for nucleon $j$. The summation runs over the two 
interacting nucleons, 1 and 2. $U$ stands for the meson-exchange $NN$ 
potential. It is, in principle, identical to the potential $V$ appearing in the
$NN$ scattering equation, except that here meson retardation effects (which are
neglected in the potential entering in Eq.(\ref{scateqn})) are kept as given 
by the Feynman prescription. Eq.(\ref{nuc_cur}) is illustrated in 
Figs.~\ref{fg2}a,b, where the contributions arising from both the positive- 
and negative-energy propagations of the intermediate nucleon are shown
explicitly. The positive-energy nucleonic current is sometimes referred to as 
the external current, while the negative-energy current is known as the pair 
diagram which, together with meson-exchange currents, constitutes what is 
called the internal current. In a relativistic formulation like the present 
one, such terminologies seem unnecessary.

The structure of the $NN\omega$ vertex, $\Gamma^\mu$ (the subscript $j=1,2$ is
omitted), required in Eq.(\ref{nuc_cur}) for the production is given by
\begin{equation}
\Gamma^\mu (p',p)
 = -i g_{NN\omega} [ F_V(p', p) \gamma^{\mu} -
 i \frac{{\kappa_\omega}}{2 m_{N}} F_T(p', p) \sigma^{\mu \nu} k_{\nu} ]   \ ,
\label{vertex}
\end{equation}
where $g_{NN\omega}$ denotes the vector coupling constant and $\kappa_\omega 
\equiv f_{NN\omega}/g_{NN\omega}$ with $f_{NN\omega}$ being the tensor coupling
constant. $p$ and $p'$ denote the incoming and outgoing nucleon four-momentum,
respectively, and $k=p-p'$ the four-momentum of the emitted $\omega$-meson. 
The functions $F_V(p', p)$ and $F_T(p', p)$ are form factors which describe 
the off-shell coupling of $\omega$ to the nucleons. They are normalized to 
unity when the $\omega$ and nucleons are on mass-shell, i.e.,
$F_{V,T}(p'{}^2=m_N^2, p^2=m_N^2, (p'-p)^2=m_{\omega}^2) = 1$. Note that in 
diagram Fig.~\ref{fg2}a we have $p'{}^2=m_N^2$, and in diagram Fig.~\ref{fg2}b, 
$p^2=m_N^2$. The produced $\omega$ is, of course, always on its mass-shell.

At present there is a considerable uncertainty in the $NN\omega$ coupling 
constants. In table 1 we have collected different sets of coupling constants 
from various analyses \cite{HPS76,GK80,fur_wat,MHE87,MMD96,ST97}. We see a 
broad range of values: $g^2_{NN\omega}/4\pi \sim 8$ to 35 for the vector 
coupling, and $\kappa_\omega \sim -0.16$ to +0.14 for the ratio of tensor to 
vector coupling. For example, in the full Bonn model \cite{MHE87} a value of
$g^2_{NN\omega}/4\pi = 20$ is required for a best fit to $NN$ data. Clearly 
this fairly large number must be considered as an effective coupling strength 
rather than as the intrinsic $NN\omega$ coupling constant. This has been shown
in a recent work by Janssen et al. \cite{Janssen}, where the contribution of 
the correlated $\pi\rho$-exchange to the $NN$ interaction has been taken into 
account explicitly. They found that the additional repulsion provided by the 
correlated $\pi\rho$-exchange allows $g^2_{NN\omega}/4\pi$ to be reduced by 
about a factor of 2, leading to an ``intrinsic'' $NN\omega$ coupling constant 
which is more in line with the value one would obtain from the SU(3) symmetry 
considerations, $g^2_{NN\omega} = 9 g^2_{NN\rho}$. In the present work, we 
adopt the vector coupling constant obtained by Janssen et al. \cite{Janssen} 
which is $g^2_{NN\omega}/4\pi = 11$. For $\kappa_\omega$, we consider the 
range of $\kappa_\omega \sim -0.3$ to $+0.3$. 

Additional uncertainties come from the vertex form factors $F_V$ and $F_T$.
Although one has some idea about the non-locality of the $NN\omega$ vertex 
from $NN$ scattering, basically nothing is known about its range relevant for 
the $\omega$-meson production process discussed here. This is because in $NN$ 
scattering the exchanged $\omega$ is off mass-shell, whereas in the present 
case the $\omega$ is produced on-mass shell and the nucleons are off their 
mass-shell. The theoretical understanding of these form factors is beyond the 
scope of the present paper. Therefore, in analogy to what is usually done in 
$NN$ potential models, we take $F_V = F_T \equiv F_N$ \cite{Mach89} in the 
present exploratory study. We assume the form factor to be of the form
\begin{equation}
F_N(l^2) = \left ( \frac{n\Lambda_N^4} {n\Lambda_N^4 + (l^2-m_N^2)^2} 
\right )^n  \ ,
\label{formfactorN}
\end{equation}
where $l^2$ denotes the four-momentum squared of either the incoming or
outgoing off-shell nucleon, $p^2$ or $p'{}^2$. The cutoff parameter $\Lambda_N$ 
and $n$ (integer) are treated as free parameters which are adjusted to fit the
$\omega$-production data. Note that if $n \rightarrow \infty$, $F_N(l^2)$ 
becomes a Gaussian function. We also introduce the form factor given by 
Eq.(\ref{formfactorN}) at those $NNM$ vertices (cf. Eq.(\ref{nuc_cur})) 
appearing next to the $\omega$-production vertex, where the (intermediate)
nucleon and the exchanged mesons are off their mass-shell. Therefore, the 
corresponding form factors are given by the product $F_N(l^2) F_{M}(q_M^2)$, 
where $M$ stands for each of the exchanged mesons. The form factor 
$F_M(q_M^2)$ accounts for the off-shellness of the exchanged meson and is 
taken consistently with the $NN$ potential used for generating the T-matrix.

Besides the nucleonic current, one might think of contributions from the 
isospin-1/2 nucleon resonances ($N^*$) to the $\omega$-emission current. 
However, there are no experimental indication of the known $N^*$ resonances 
decaying into $\omega + p$. These isospin-1/2 and other nucleon resonances can,
therefore, contribute to the $\omega$-meson production only via coupled 
channels. Since in the present work we restrict ourselves to the energy region
far below nucleon resonance threshold in the final state, the coupling to such
a channel should not induce any significant effects. In particular, we do not 
expect any significant modification of the energy dependence of the 
$\omega$-production cross section due to such a coupling.

For the meson-exchange current, $J^\mu_{mec}$, we consider the contribution
from the $\pi\rho\omega$ vertex which gives rise to the dominant 
meson-exchange current (Fig.~\ref{fg2}c). The $\pi\rho\omega$ vertex required
for constructing the meson-exchange current is derived from the Lagrangian 
density
\begin{equation}
{\cal L}_{\pi\rho\omega} = \frac{g_{\pi\rho\omega}} {m_\omega}
\varepsilon_{\alpha\beta\nu\mu} \partial^\alpha \vec \rho^\beta(x) \cdot
\partial^\nu \vec \pi(x) \omega^\mu(x) \ ,
\label{pirhoomega}
\end{equation}
where $\varepsilon_{\alpha\beta\nu\mu}$ denotes the Levi-Civita antisymmetric
tensor with $\varepsilon_{0123}=-1$. The exchange current is then given by
\begin{equation}
J^\mu_{mec} = [\Gamma^\alpha_{NN\rho}(q_\rho)]_1 iD_{\alpha\beta}(q_\rho)
              \Gamma^{\beta\mu}_{\rho\pi\omega}(q_\rho, q_\pi, k_\omega)
              i\Delta(q_\pi) [\Gamma_{NN\pi}(q_\pi)]_2   +  (1 \leftrightarrow 2) \ ,
\label{mec_cur}
\end{equation}
where $D_{\alpha\beta}(q_\rho)$ and $\Delta(q_\pi)$ stand for the $\rho$- and 
$\pi$-meson (Feynman) propagators, respectively. The vertices $\Gamma$ 
involved are self-explanatory.

The coupling constant $g_{\pi\rho\omega}$ can be estimated from the decay of
the $\omega$-meson into $\pi^o$ and $\gamma$ in conjunction with vector-meson 
dominance.  According to  Ref.\cite{Durso} $g_{\pi\rho\omega}= 10$ at 
vanishing four-momentum square of the $\rho$-meson. The sign of 
$g_{\pi\rho\omega}$, which determines the relative sign between the nucleonic 
and meson-exchange currents, can be inferred, for example, from the analysis of
pion photo-production off the nucleon in the $1$ GeV region \cite{sign}.

Each vertex in Eq.(\ref{mec_cur}) is accompanied by a form factor. For the 
$NN\pi$ vertex we use a monopole form factor $F_\pi(q_\pi^2) = 
(\Lambda_\pi^2 - m_\pi^2) / (\Lambda_\pi^2 - q_\pi^2) $ with $\Lambda_\pi = 
1000$ MeV. Note that the cutoff parameter $\Lambda _\pi$ required in 
meson-exchange models of the $NN$ interaction is usually larger than what we 
use here (e.g., $\Lambda_\pi = 1700$ MeV in the NN model applied in the present
paper). However, as has been also shown in the previously mentioned work of 
Janssen et al. \cite{Janssen}, the $NN\pi$ vertex parameters used in 
conventional meson-exchange models must be likewise considered as effective 
values, parametrizing, among other effects, (missing) contributions from the 
correlated $\pi\rho$-exchange. Indeed, once these contributions are taken
into account explicitly (cf. Ref.\cite{Janssen}) the cutoff parameter of the 
$NN\pi$ vertex goes down to about $\Lambda_\pi=1000$ MeV which is closer to 
the value of $\Lambda_\pi \sim 800$ MeV obtained from other sources
\cite{Coon,ThHo,lattice}. For the $NN\rho$ vertex, we use $F_\rho(q_\rho^2) =
((\Lambda_\rho^2 - m_\rho^2) / (\Lambda_\rho^2 - q_\rho^2))^2 $ with
$\Lambda_\rho = 1850$ MeV, consistent with the value in the $NN$ potential 
\cite{Mach89} that has been used to generate our T-matrix. Since nothing is
known about the form factor at the $\pi\rho\omega$ vertex where both the
$\pi$ and $\rho$-meson are off their mass-shell, we assume the form
\begin{equation}
F_{\pi\rho\omega}(q_\pi^2, q_\rho^2) = F_\pi(q_\pi^2)
\left ( \frac{\Lambda_\rho^2} {\Lambda_\rho^2 - q_\rho^2} \right )^2 \ ,
\label{formfactorM}
\end{equation}
with $F_\pi(q_\pi^2)$ given above and $\Lambda_\rho = 1850$ MeV. It is 
normalized to unity at $q_\pi^2 = m_\pi^2$ and $q_\rho^2 = 0$ consistent with 
the kinematics at which the value of the coupling constant $g_{\pi\rho\omega} =
10$ was determined. Quite recently, Friman and Soyeur \cite{Friman} have used 
the interaction Lagrangian given by Eq.(\ref{pirhoomega}) in conjunction with 
the vector-dominance model and a form factor of the form $\Lambda_\rho^2 / 
(\Lambda_\rho^2 - q_\rho^2)$ with $\Lambda_\rho = m_\rho$ at the 
$\pi\rho\omega$ vertex in order to fit the Dalitz decay of the $\omega$-meson 
into $\pi^o$ and $\mu^+ \mu^-$. Although this is in a quite different kinematic
regime from that involved in the present meson production reaction, we see that
our choice of the $\pi\rho\omega$ form factor given by Eq.(\ref{formfactorM}) 
would lead to an equivalent monopole form factor with $\Lambda_\rho \sim 1300$
MeV when the pion is on its mass-shell.

There are, in principle, other meson-exchange currents that could contribute
to the $\omega$-meson production in $pp$ collisions. One such potential
candidate is due to the $\eta\omega\omega$ vertex which can be obtained from a
Lagrangian analogous to that of Eq.(\ref{pirhoomega}). The corresponding
coupling constant can be estimated from the decay width of $\omega$ into
$\eta$ and $\gamma$ assuming vector dominance. This yields 
$g_{\eta\omega\omega} \cong 7$, which is comparable to the $\pi\rho\omega$ 
coupling constant $g_{\pi\rho\omega} \cong 10$. However, the meson-exchange 
current due to the $\eta\omega\omega$ vertex is about two orders of magnitude 
smaller than the one involving the $\pi\rho\omega$ vertex, the main reason 
being the smallness of the $NN\eta$ and $NN\omega$ (tensor) couplings compared
to the $NN\pi$ and $NN\rho$ couplings, respectively. The $\eta$-meson 
propagator reduces the cross section only by a factor of 2 compared to the 
pion propagator. Another exchange current may be due to the 
$\sigma\omega\omega$ vertex, whose coupling constant may be estimated from the
decay width of $\omega$ into $\gamma$ and $\pi^+\pi^-$ or $\pi^0\pi^0$. 
Assuming an interaction Lagrangian analogous to that given by 
Eq.(\ref{pirhoomega}), we find $g_{\sigma\omega\omega} \sim 0.5$, which is 
extremely small.

The $\omega$-emission current defined above is, in part, just the Born term of
a more general current which can be obtained by using the T-matrix amplitude 
for the $M + p \rightarrow \omega + p$ transition, where $M$ denotes any meson
of interest. In fact, if we disregard the nucleon labelled 2 in Fig.~\ref{fg2} 
the current becomes nothing else than the Born term of the $M + p \rightarrow 
\omega + p$ transition amplitude. Diagram \ref{fg2}a, then, would correspond to
the s-channel process referred to as the direct pole term, while diagram 
\ref{fg2}b would correspond to the u-channel process referred to as the 
exchange (or cross) pole term. Diagram \ref{fg2}c would correspond to the 
pion-exchange t-channel process. The T-matrix amplitude for the $M + p 
\rightarrow \omega + p$ process may be separated into the so called pole and 
non-pole terms a la Pearce and Afnan \cite{Afnan}. The pole term is defined 
just as the (Born) direct pole term mentioned above with the physical nucleon 
mass and physical $NN\omega$ and $NNM$ vertices. The non-pole term is, then, 
the difference between the full T-matrix and its pole term. Since in the 
nucleonic current given by Eq.(\ref{nuc_cur}) we use the physical nucleon mass
and physical $NN\omega$ and $NNM$ vertices, the pole term of the T-matrix 
amplitude is fully accounted for in the present work. What is taken in the Born
approximation is, therefore, the non-pole part of the T-matrix only. 

 \section{Results}

\begin{figure}
\epsfig{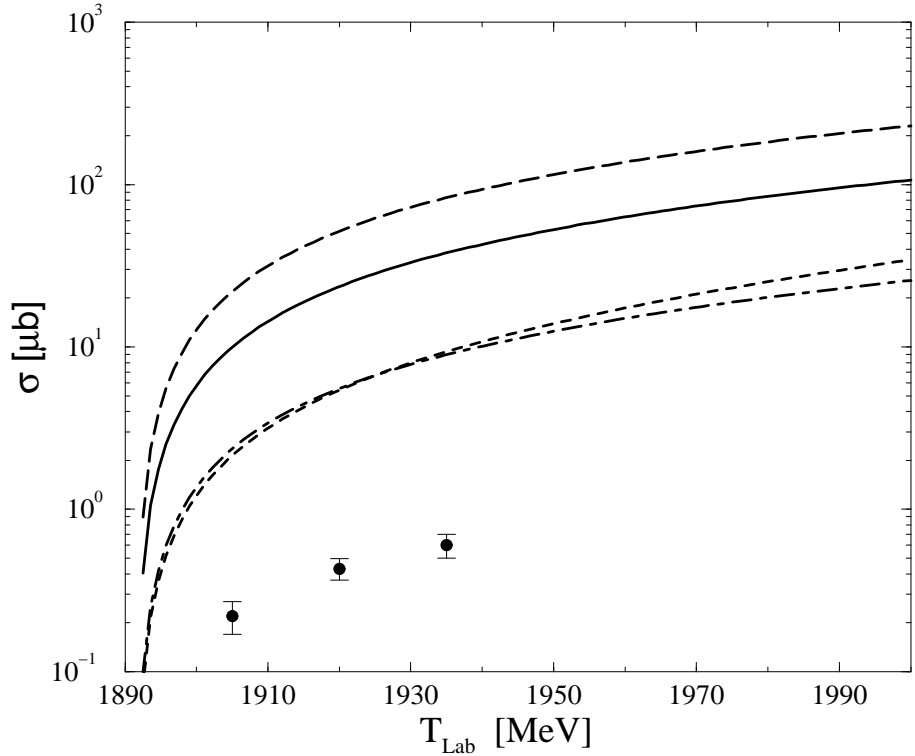}
\caption{Total cross section for the reaction $p + p \rightarrow 
p + p + \omega$ as a function of incident energy. The theoretical results 
shown here were obtained without including form factors at the 
$\omega$-production vertices. The short-dashed line is the contribution of the
positive-energy nucleonic current alone, the long-dashed line is the 
contribution of the total (positive + negative) nucleonic current, and the 
dash-dotted line is the contribution of the meson exchange current. The solid 
line is the coherent sum of the all contributions. The experimental data are 
from Ref.\protect\cite{Saclay}.}
\label{fg3}
\end{figure}

Once all the ingredients are specified, the total cross section for the 
reaction $p + p \rightarrow p + p + \omega$ can be calculated. Let us first 
consider the case where no form factors are used at the $\omega$ production 
vertices (see Eqs.(\ref{vertex},\ref{mec_cur})). We found that the effect of 
the tensor coupling is essentially to change the absolute magnitude of the 
cross section without affecting its shape as a function of incident energy in 
the energy domain considered in this work. The value of $\kappa_\omega = -0.3$
yields the largest cross section in the range considered for $\kappa_\omega$,
and that the choice of $\kappa_\omega=+0.3$ would lead to a reduction of the 
cross section by about a factor of 2. In what follows, we assume 
$\kappa_\omega = -0.3$. Corresponding results are shown in Fig.~\ref{fg3} as a
function of the incident proton laboratory energy, together with preliminary 
data of the SPES3 collaboration at Saclay \cite{Saclay}. We see that the 
contributions from the nucleonic (long-dashed line) as well as the mesonic 
(dash-dotted line) current are much larger than the experimental results. 
Although these contributions interfere destructively, the total result (solid 
line) still overestimates the data by nearly two orders of magnitude. Note 
that in Ref.\cite{Saclay} both the original data and the data shifted in energy
(based on a fit to the phase-space energy dependence) have been presented. We 
show here only the nominal data as the disagreement of the nominal data with 
the phase-space behaviour in Ref.\cite{Saclay} can be due to dynamical effects
calculated explicitly in the present paper.

Let us take a closer look at the contribution from the nucleonic current. In 
the following, we refer to the parts of the nucleonic current arising from the
positive-(negative-) energy nucleon propagation in Eq.(\ref{nuc_cur}) as the 
positive-(negative-) energy nucleonic current. As can be seen from 
Fig.~\ref{fg3}, the large nucleonic current contribution (long-dashed line) is
dominantly due to the negative-energy component. The positive-energy nucleonic
current contribution (short-dashed line) is relatively small. This is opposite
to what is known from the $p + p \rightarrow p + p + \gamma$ ($pp\gamma$) 
reaction. There, both the negative-energy nucleonic and meson-exchange 
currents are higher-order corrections to the dominant positive-energy 
nucleonic current \cite{Kanzo2}. The role of the positive- and negative-energy
nucleonic currents can be most easily understood if we consider the $\omega$ 
production at threshold. We focus on the contribution from the pre-emission 
diagrams (Fig.~\ref{fg2}b) which dominate over the post-emission diagrams 
(Fig.~\ref{fg2}a) at threshold. The transition amplitude given by 
Eq.(\ref{amplitude}) may then be expressed as
\begin{eqnarray}
M &\equiv & < \vec p\ ', S'M_S'| \epsilon^{*\mu} M_\mu | \vec p, SM_S > 
\nonumber \\
  &\sim  & \left ( {m_N\over \varepsilon(\vec p)} \right ) \sum_{S''M_S''}
\Bigg \{
\frac{< \vec p\ ', S'M_S'| T^+ | \vec p, S''M_S'' >
 < S''M_S'' | \epsilon^{*\mu} \Gamma^+_\mu | SM_S>} {- m_\omega}  \nonumber \\
  &+      & \frac{< \vec p\ ', S'M_S'| T^- | \vec p, S''M_S'' >
 < S''M_S'' | \epsilon^{*\mu} \Gamma^-_\mu | SM_S>} {2\varepsilon(\vec p) - 
m_\omega} \Bigg \}      \ ,
\label{nuc_ampl_thre}
\end{eqnarray}
where $\vec p$ ($\vec p\ '$) denotes the initial (final) relative momentum of 
the two interacting nucleons, $S$ ($S'$) and $M_S$ ($M_S'$) stand for the 
inital (final) total spin and its projection. $\epsilon^\mu$ denotes the 
polarization vector of the emitted $\omega$-meson. Here, the superscript 
($+/-$) in $T$ and $\Gamma_\mu$ denotes terms involving a positive- or
negative-energy nucleon. $\varepsilon(\vec p) = \sqrt{\vec p\ ^2 + m_N^2}$,
and $\vec p\ ' = 0$ at threshold. In the above equation, 
$\Gamma^\pm_\mu \equiv \Gamma^\pm_{1 \mu} + \Gamma^\pm_{2\mu}$, cf. 
Eq.(\ref{nuc_cur}). Its matrix elements are given by
\begin{multline} < S''M_S'' | \epsilon^{*\mu} \Gamma^+_\mu | SM_S> = g_{NN\omega}
\frac{\kappa_\omega} {4m_N} \left ( {m_\omega \over m_N} \right ) |\vec p|
[1 - (-)^{S''-S}] \\ \times < S''M_S''| \vec \sigma_1 \cdot (\hat p \times 
\vec \epsilon) | SM_S >  \ ,
\label{Gamma_12_pos}
\end{multline}
with $\hat p = \vec p / |\vec p|$, and
%
\begin{multline} \label{Gamma_12_neg}
< S''M_S''| \epsilon^{*\mu} \Gamma^-_\mu | SM_S>= 
i g_{NN\omega} [1 + (-)^{S''-S}] \\
 \times  \Bigg \{
\left ( 1 - \kappa_\omega{m_\omega \over 2m_N} \right )
\left ( {\varepsilon(\vec p) - m_N \over m_N} \right ) \vec \epsilon \cdot 
\hat p < S''M_S''| \vec \sigma_1 \cdot \hat p | SM_S > \\
 -  \left ( {\varepsilon(\vec p)\over m_N} +
               \kappa_\omega{m_\omega \over 2m_N} \right )
       < S''M_S''| \vec \sigma_1 \cdot \vec \epsilon | SM_S > \Bigg \}  \ .
\end{multline}
%

The first term in Eq.(\ref{nuc_ampl_thre}) corresponds to the positive-energy
and the second term to the negative-energy nucleonic current contribution. It
is obvious that, due to the non-zero mass of the $\omega$-meson, the ratio
between the negative- and positive-energy nucleonic current contributions
is much larger compared to the $pp\gamma$ reaction. This is simply a 
consequence of the fact that the intermediate nucleon is far off-shell at 
least by an amount of the mass of the emitted $\omega$-meson. In the 
$pp\gamma$ reaction, the positive-energy contribution leads to the well known 
infrared divergence due to the massless nature of the emitted photon. In the 
present case, a further enhancement of the negative-energy contribution 
relative to the one from the positive-energy current arises from the smallness
of the tensor-to-vector coupling ratio, $\kappa_\omega$, as can easily be seen
by comparing Eqs.(\ref{Gamma_12_pos}) and (\ref{Gamma_12_neg}).

It is also interesting to note that the positive-energy nucleonic current
contribution at threshold involves only the tensor coupling $f_{NN\omega}=
g_{NN\omega}\kappa_\omega$ (cf. Eq.(\ref{Gamma_12_pos})). This can be 
understood as follows: in the limit of small momentum of the emitted 
$\omega$-meson, i.e., $\vec k \rightarrow 0$, the matrix element of the 
$NN\omega$ vertex for the $j-th$ nucleon is given by 
\begin{equation}
<\bar u(\vec p) | \epsilon^*_\mu \Gamma_j^\mu |
u(\vec p) > = -i\vec \epsilon \cdot [ g_{NN\omega}(\vec p_j / m_N) +
i (f_{NN\omega}/4m_N)(m_\omega/m_N)(\vec \sigma_j \times \vec p_j) ].
\label{NNvertex1}
\end{equation}
At threshold, the Pauli principle requires the two protons to be in a 
relative $^{1}S_0$ final state which, together with the $\omega$-meson having 
total $J^P=1^-$, implies that the initial two protons are in a $^{3}P_1$ 
state. To get a non-vanishing contribution the production operator therefore 
has to change the total spin of the $NN$ subsystem which is possible only with 
the tensor coupling and not with the vector coupling as can be seen from the 
expression for the $\omega$-production vertex given by Eq.(\ref{NNvertex1}). 
Actually, the feature that the positive-energy nucleonic current contribution 
is independent of the vector coupling, $g_{NN\omega}$, holds not only at 
threshold but also at any incident energy, provided the momentum of the 
emitted $\omega$-meson goes to zero. This is because in this limit the vector 
coupling leads only to the convection current for positive-energy nucleon 
propagation (first term in Eq.(\ref{NNvertex1})) and the total convection 
current should vanish for identical particles in their center-of-mas frame. 

In principle, there are two ways for bringing the theoretical results in 
agreement with the experimental data. One is simply to readjust the relevant 
coupling constants in order to obtain a more complete cancellation between the
nucleonic and mesonic currents. The other possibility is to introduce form 
factors at the $\omega$-production vertices, as discussed in section II. The 
first alternative would require a rather drastic change of the coupling 
constants, beyond the uncertainties discussed in the previous section. 
Therefore, in the present analysis we choose the second alternative. The 
introduction of form factors within our model is also well motivated because 
the $\omega$-meson production reaction is a highly off-shell process. Indeed, 
in both the nucleonic and meson-exchange current diagrams shown in 
Fig.~\ref{fg2}a-c, all the intermediate particles are highly off their 
mass-shell. Specifically, for an incident energy corresponding to the 
$\omega$-production threshold, the intermediate nucleon in diagram 
Fig.~\ref{fg2}a is off its mass-shell by $p^2 - m_N^2 \sim 2.3$ GeV$^2$ and 
the nucleon in diagram Fig.~\ref{fg2}b is $p^2 - m_N^2 \sim -1.5$ GeV$^2$ 
off-shell. Similarly, for the $\pi$- and $\rho$-mesons in diagram 
Fig.~\ref{fg2}c we have $q_\pi^2 - m_\pi^2 \sim  -0.8$ GeV$^2$ and 
$q_\rho^2 - m_\rho^2 \sim -1.4$ GeV$^2$.

\begin{figure}
\epsfig{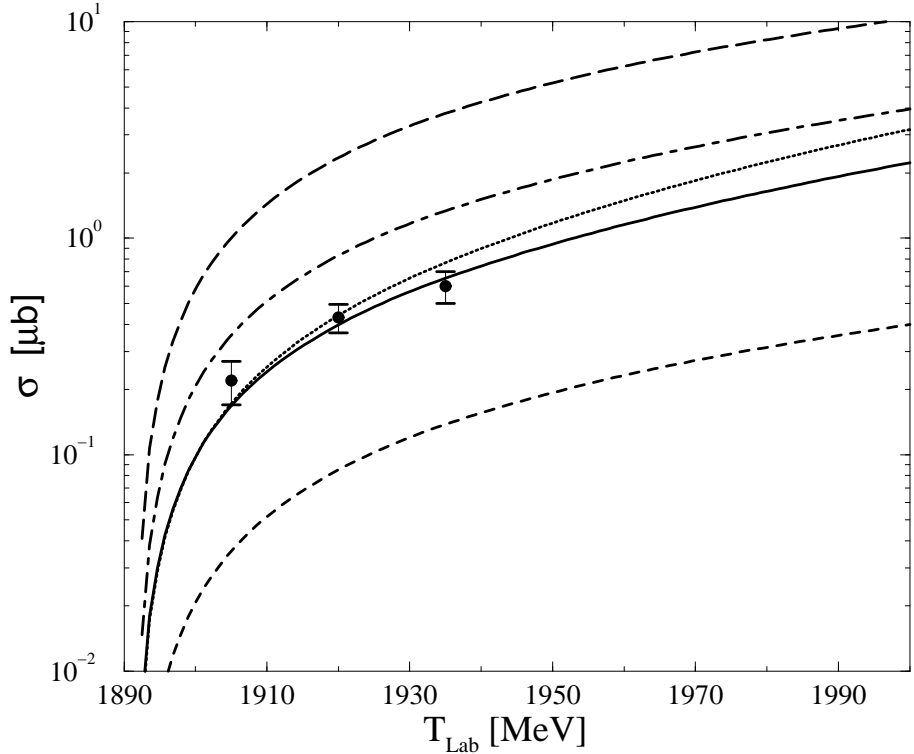}
\caption{Total cross section for the reaction $p + p \rightarrow 
p + p + \omega$ as a function of incident energy. The theoretical results 
shown here include form factors as described in Section II. The dash-dotted 
line is the contribution of the meson exchange current. The long-dashed 
(short-dashed) line is the contribution of the nucleonic current based on the 
cutoff mass $\Lambda_N$ = 1160 (850) MeV (cf. Eq.(5)). The coherent sum of the
two contributions where the nucleonic current is larger than the mesonic 
current ($\Lambda_N$ = 1160 MeV) is given by the dotted line, the one where 
the nucleonic current is smaller than the mesonic current ($\Lambda_N$ = 850 
MeV) is given by the full line. The experimental data are from 
Ref.\protect\cite{Saclay}.}
\label{fg4}
\end{figure}

Results including the form factors (Eqs. (\ref{formfactorN},\ref{formfactorM}))
at the $\omega$ production vertices (Eqs. (\ref{vertex},\ref{mec_cur})) are 
shown in Fig.~\ref{fg4}. With our choice for the $\pi\rho\omega$ vertex (cf. 
the discussion after Eq.(\ref{formfactorM})) the cutoff mass $\Lambda_N$ is 
the only free parameter to be adjusted to the data. (The parameter $n$ in 
Eq.(\ref{formfactorN}) has been fixed to be $n=1$.) Since the meson-exchange 
current contribution alone already exceeds the data (cf. dash-dotted line in 
Fig.~\ref{fg4}) and due to the destructive interference between the nucleonic 
and mesonic currents we can find two solutions, one with the nucleonic current
contribution being larger than the mesonic current contribution, and another 
one with the nucleonic current being smaller than the mesonic current 
contribution. The long-dashed line is the nucleonic current contribution
corresponding to $\Lambda_N=1160$ MeV in Eq.(\ref{formfactorN}) which yields 
the total contribution given by the dotted line ($NC > MEC$) when it is added 
coherently to the mesonic current (dash-dotted curve). The short-dashed line 
is the nucleonic current contribution corresponding to $\Lambda_N=850$ MeV 
which leads to the total contribution given by the solid line ($NC < MEC$). We
see that the energy dependence of the two solutions are different. Since the 
form factors necessarily introduce an energy dependence, we have checked 
whether this difference in the energy dependence between the two solutions is 
just an artifact of the particular form of the form factor 
(Eq.(\ref{formfactorN})) used in the present work. To this end, we employed 
different types of form factors at the $NN\omega$ vertex, such as a Gaussian 
form. We found that the feature exhibited by the two solutions in 
Fig.~\ref{fg4} persists.

\begin{figure}
\epsfig{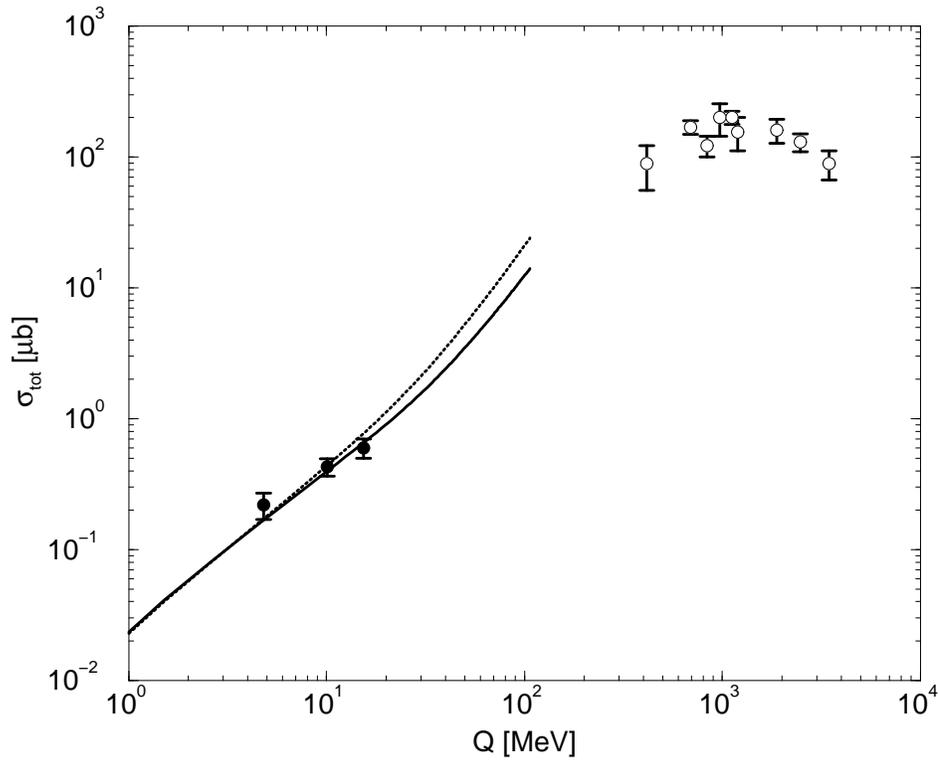}
\caption{Total cross section for the reaction $p + p \rightarrow 
p + p + \omega$ as a function of the excess energy $Q=\sqrt{s}-\sqrt{s_o}$. 
The curves are the same as in Fig.~\protect\ref{fg4} extended up to $T_{lab} =
2.2\ GeV$. The high energy data are taken from Ref.\protect\cite{Flam}.}
\label{fg5}
\end{figure}

In order to demonstrate the difference in the energy dependence our results 
are shown again in Fig.~\ref{fg5} over a larger energy range and as a function
of excess energy, $Q \equiv \sqrt{s} - \sqrt{s_o}$, where $\sqrt{s}$ denotes 
the total energy of the system and $\sqrt{s_o}=2m_N + m_\omega$, its  
$\omega$-production threshold energy. Here we also display the experimental 
data from Ref.\cite{Flam} at higher energies. Our model calculations are 
carried out up to the energy at which the FSI becomes inelastic, which 
corresponds to an incident energy of around $T_{lab} = 2.2$ GeV. Already in 
this energy range one observes a difference in the predicted cross sections 
for the two scenarios ($NC {< \atop >} MEC$) considered. Therefore, in 
principle, measurements of the cross section for $Q \sim 100$ MeV and higher 
could be useful in determining the relative magnitude of the two production 
mechanisms.

\begin{figure}
\epsfig{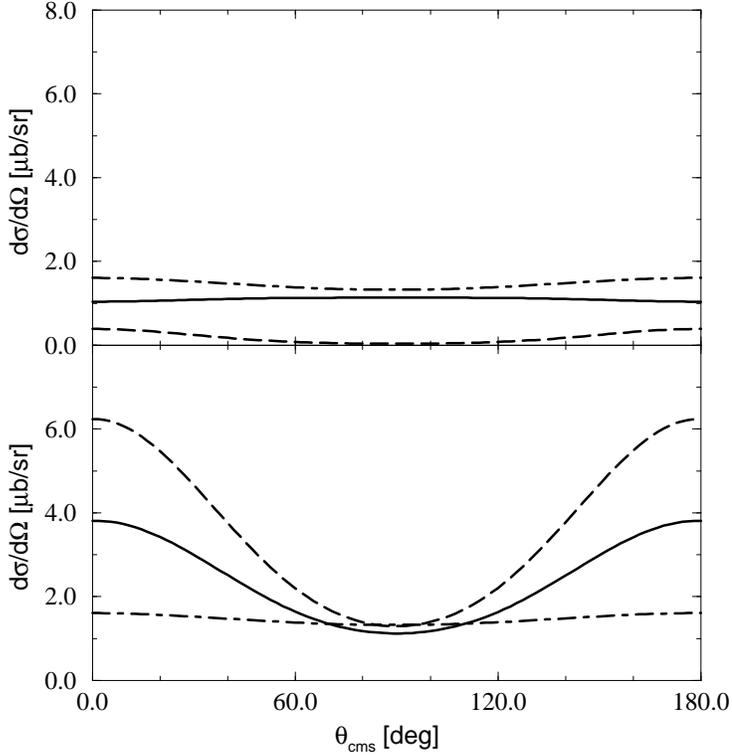}
\caption{Angular distribution of the emitted $\omega$ mesons in the total CM 
system at the energy $T_{lab}= 2.2\ GeV$. The lower figure shows the result
where the nucleonic current is larger than the mesonic current ($NC > MEC$)
whereas the upper graph contains the results where the nucleonic current is
smaller than the mesonic current ($NC < MEC$). The dash-dotted line is the 
contribution of the mesonic current, the long-dashed line is the contribution 
of the nucleonic current and the full line is the total result.}
\label{fg6}
\end{figure}

In Fig.~\ref{fg6} we present angular distributions of the emitted $\omega$ 
mesons in the total center-of-mass (CM) system at an incident energy of 
$T_{lab}=2.2$ GeV for the two scenarios discussed above ($NC {< \atop >}MEC$).
As can be seen, the two scenarios lead to dramatically different angular 
distributions: strong anisotropy in the case of nucleonic current contribution
being larger than the mesonic current contribution ($NC > MEC$, lower figure) 
and an almost isotropic distribution in the case of nucleonic current 
being smaller than the mesonic current contribution ($NC < MEC$, upper 
figure). The strong anisotropy of the $cos^2\theta$ shape introduced by the 
nucleonic current is due to the spin-dependent part of the current, sometimes 
referred to as the magnetization current \cite{magnet}. To leading order, its 
contribution to the differential cross section is given by
\begin{equation}
{d^2\sigma_{magn} \over dW d\Omega} \cong \left ( a^2 + b^2v'{}^2 \right ) +
 c^2 v^2 cos^2(\theta)     \ ,
\label{magnxsc}
\end{equation}
where $v$ ($v'$) denotes the relative velocity of the two interacting nucleons
in the initial (final) state. $a$, $b$, and $c$ are smooth functions of the
energy $W$ of the emitted $\omega$-meson and of the total energy of the system.
It should be clear from Fig.~\ref{fg6} that the angular distribution provides
an unique and clear signal for discriminating between the two considered
$\omega$-production scenarios. We also note that if the mesonic current
contribution were larger than the present prediction, we would have a much more
pronounced angular dependence for the case of $NC < MEC$ than that shown in
Fig.~\ref{fg4}. Indeed, the angular distribution would exhibit a peak around
$\theta = 90^0$, where it shows a valley in the case of $NC > MEC$. We 
emphasize that the shape of the angular distribution is determined by the 
magnitude of each contribution. Therefore, by measuring the angular 
distribution one should be able to extract uniquely the magnitude of the 
individual contributions.

\begin{figure}
\epsfig{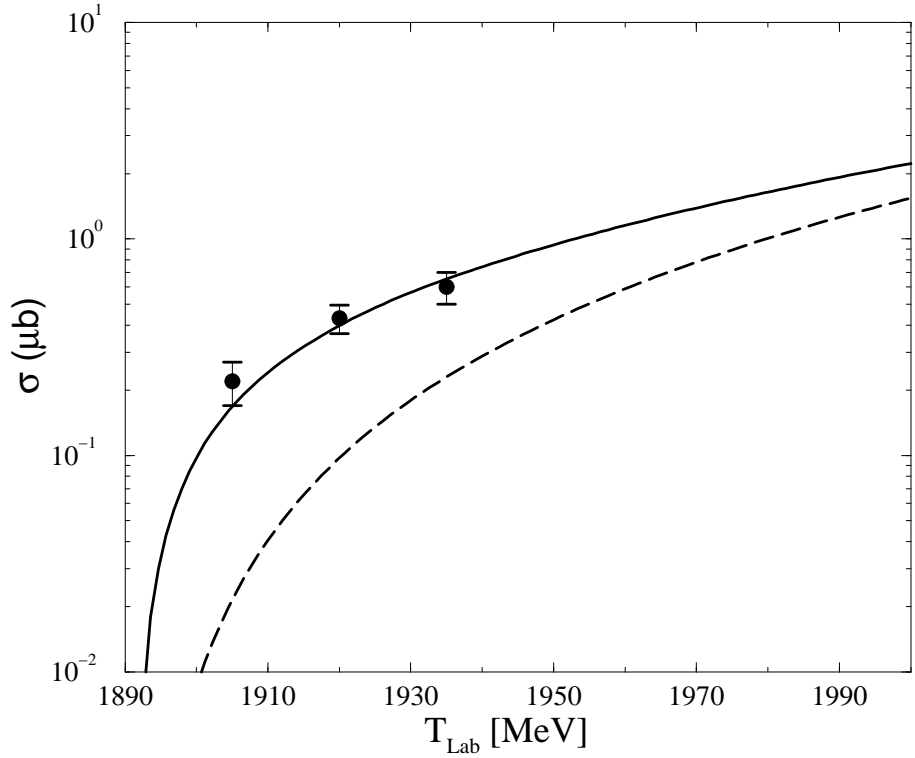}
\caption{The effect of the final state interaction on the total 
$\omega$-production cross section. The full and dashed lines correspond to the
results with and without final state interaction, respectively. These results
correspond to the choice of the nucleonic current being larger than the mesonic
current ($NC > MEC$). }
\label{fg7}
\end{figure}

The effect of the FSI is extremely important in any particle production 
process where the interacting nucleons are left in a low-energy S-wave state. 
This is demonstrated in Fig.~\ref{fg7} where the results with and without FSI 
are shown. Close to the threshold the FSI enhances the total cross section by 
almost an order of magnitude. No FSI effects were taken into account in recent
calculations \cite{Sib96,CLK97}. In view of this, in our opinion, the rather 
good agreement of the peripheral on-shell rescattering model \cite{Sib96} must
be considered as rather accidental.

\section{Summary}

We have investigated the $p + p \rightarrow p + p + \omega$ reaction within
a relativistic meson-exchange model. It has been found that the nucleonic and
$\pi\rho\omega$ exchange currents are the two potential sources contributing to
the $\omega$-production in this reaction, and that, they interfere 
destructively. The calculation ignoring the non-localities of the $\omega$ 
production vertices leads to a dramatic overestimation of preliminary SPES3 
data \cite{Saclay}. We interprete this as a manifestation of off-shell effects
in hadronic vertices within our model. In the absence of a microscopic 
prescription, we introduce phenomenological form factors to account for the 
off-shell effects in the vertex functions and fix the free parameters by 
fitting to the SPES3 data. It is, then, found that the positive-energy 
nucleonic current contribution is very small compared to the negative-energy 
nucleonic and the $\pi\rho\omega$ meson-exchange current contribution.

We find that the relative magnitude of the contributions from the nucleonic
current and the meson-exchange currents influences the energy dependence of 
the predicted $p + p \rightarrow p + p + \omega$ cross section. Therefore, 
from a measurement of the cross sections as a function of incident energy one 
should, in principle, be able to identify the dominant reaction mechanism for 
$\omega$ production. As our most important result, we find that the angular 
distribution of the emitted $\omega$-meson depends sensitively on the strength
of the individual contributions. A study of the angular distribution should 
therefore allow one to determine uniquely the magnitude of each contribution. 
A measurement of this observable, which can be carried out at modern 
accerelators, would be of great importance for understanding $\omega$ 
production in $NN$ collisions.

\hskip 2cm

{\bf Acknowledgments}
We are indebted to Wolfgang K\"uhn for a discussion of experiments being 
carried out at SATURNE and to Collin Wilkin and Francois Hibou for exchange of
information concerning the SPES3 experiment \cite{Saclay}. We also thank Gary
Love for a careful reading of the manuscript. This work was partially 
supported by the DLR grant, project no. POL-81-94.

\vfill \eject

\vfill \eject

\begin{table}

\caption{A compilation of $NN\omega$ coupling constants.}

\begin{center}
\begin{tabular}{|c|c|c|c|c|c|c|}
     $g^2 / 4 \pi$ & $f^2 / 4 \pi$ &  $g$   &  $f$  & $f/g$  &   Ref.        &
    method \\
\hline

  23.03         & 0.33   &  17.01      & -2.04        & -0.12              &
               & naive VDM \\
  24 $\pm$ 12     &   $<$ 1. &  17.37      &              &               & 
\cite{HPS76}  & EM form factors \\
   8.1$\pm$ 1.5  &0.16$\pm$0.45 & 10.09   &  1.42        &  0.14$\pm$0.20 & 
\cite{GK80}   & NN forward d.r. \\
  35.41         & 0.90       &   21.094    & -3.37        & -0.16         & 
\cite{fur_wat}& EM form factors \\
  20.           &  0.0   &   15.85     &   0.0        &  0.0          & 
\cite{MHE87}  & NN scattering  \\
  34.6$\pm$ 0.8   & 0.93   &20.86$\pm$ 0.25&-3.41 $\pm$ 0.24& -0.16 $\pm$ 0.01
& \cite{MMD96}  & EM form factors \\
              &        &             &  -2 to 2     &               & 
\cite{ST97}   & $\omega\pi\pi\pi$ anomaly \\
  11          &        &   11.8      &              &               & 
\cite{Janssen} & NN scattering \\
\hline
\end{tabular}\end{center}

\end{table}

\end{document}